\documentclass[prd,amsmath,amssymb,showpacs,superscriptaddress,nofootinbib,twocolumn]{revtex4-1}
\usepackage{graphicx}
\usepackage{epsfig,graphics,subfigure,psfrag,amsmath,amssymb}
\usepackage{lineno}
\usepackage{dcolumn}
\usepackage{bm}
\usepackage{overpic}
\usepackage{xspace}
\usepackage{xcolor}
\usepackage{rotating}
\usepackage{color}
\usepackage{multirow}
\usepackage{colortbl}
\usepackage{epstopdf}
\usepackage{float}
\usepackage{makecell}
\usepackage{mathrsfs} 
\usepackage{breqn}  
\usepackage{tabularx}
\usepackage{booktabs}
\usepackage{enumitem}
\usepackage[compat=1.1.0]{tikz-feynman}
\usepackage{tikz-feynhand}
\usepackage[colorlinks,linkcolor=blue,anchorcolor=blue,citecolor=blue]{hyperref}
\newcommand{\gev}{\ensuremath{\mathrm{\,Ge\kern -0.1em V}}\xspace}
\newcommand{\mev}{\ensuremath{\mathrm{\,Me\kern -0.1em V}}\xspace}
\newcommand{\mevcc}{\ensuremath{{\mathrm{\,Me\kern -0.1em V\!/}c^2}}\xspace}

\def\fz#1       {\ensuremath{f_0({#1})}\xspace}

\usepackage{lineno}
\usepackage{hyperref}
\hypersetup{colorlinks = true,
            linkcolor = black,
            urlcolor = blue,
            citecolor = blue}

\newcommand{\BESIIIorcid}[1]{\href{https://orcid.org/#1}{\hspace*{0.1em}\raisebox{-0.45ex}{\includegraphics[width=1em]{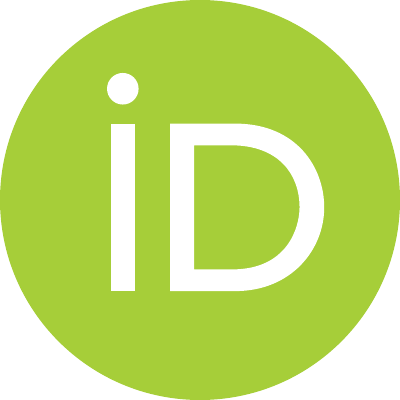}}}} 
            
\begin{document}

\title{\bf \boldmath Measurement of the double Dalitz decay $\eta \to e^+e^-e^+e^-$}
\author{BESIII Collaboration}
\thanks{Full author list given at the end of the paper}

\begin{abstract}
Using a data sample of $(1.0087 \pm 0.0044) \times {10^{10}}$ $J/{\psi}$ events collected with the BESIII detector, we study the rare double Dalitz decay of $\eta\rightarrow e^+e^-e^+e^-$ through the processes $J/\psi\rightarrow \gamma \eta$ and $J/\psi\rightarrow \gamma \eta' ,\eta' \to \pi^+\pi^-\eta$. Clear $\eta$ signals are observed in the $e^+e^-e^+e^-$ invariant mass spectrum, with statistical significances of 5.9$\sigma$ and 7.8$\sigma$ for the two channels, respectively. By combining both modes, we determine the branching fraction of $\eta\rightarrow e^+ e^- e^+ e^-$ to be $(2.63~\pm~0.34_{\rm stat}~\pm~0.16_{\rm syst}) \times10^{-5}$. The result is consistent with the previous measurements within uncertainties and further constrains physics beyond the standard model. 
\end{abstract}

\maketitle

\section{INTRODUCTION}
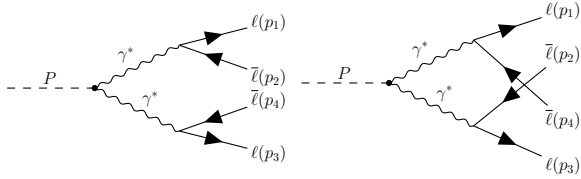
\begin{figure}[h]
\centering
\hspace*{-0.3cm}
\begin{minipage}[b]{0.46\linewidth}
\begin{tikzpicture}[scale=0.45, transform shape, baseline=(current bounding box.center)]
  \begin{feynman}
    \vertex (a) at (0,0);
    \vertex[dot] (b) [right=of a] at (1.0,0){};
    \vertex [above right=of b] (c) at (4.0, 0.2);
    \vertex [below right=of b] (d) at (4.0, -0.2);
    \vertex [above right=of c] (f1) at (6, 0.5) {\Large ${\ell}(p_1)$};
    \vertex [below right=of c] (f2) at (6, 1.8) {\Large $\overline{\ell}(p_2)$};
    \vertex [above right=of d] (f3) at (6, -1.8) {\Large $\overline{\ell}(p_4)$};
    \vertex [below right=of d] (f4) at (6, -0.5) {\Large ${\ell}(p_3)$};
    \diagram* {
      (a)--[scalar, edge label={\Large $P$}](b),
      (b)--[photon, edge label={\Large $\gamma^*$}](c),
      (b)--[photon, edge label={\Large $\gamma^*$}](d),
      (f2)--[fermion](c)--[fermion](f1),
      (f3)--[fermion](d)--[fermion](f4),
    };
  \end{feynman}
\end{tikzpicture}
\end{minipage}
\hspace{-0.3cm} 
\begin{minipage}[b]{0.46\linewidth}
\begin{tikzpicture}[scale=0.45, transform shape, baseline=(current bounding box.center)]
  \begin{feynman}
    \vertex (o) at (10,0);
    \vertex[dot] (x) [right=of o] at (11.0,0){};
    \vertex [above right=of x] (y) at (14.0, 0.2);
    \vertex [below right=of x] (z) at (14.0, -0.2);
    \vertex [above right=of y] (f5) at (16, 0.7) {\Large ${\ell}(p_1)$};
    \vertex [below right=of y] (f6) at (16, 0.4) {\Large $\overline{\ell}(p_4)$};
    \vertex [above right=of z] (f7) at (16, -0.6) {\Large $\overline{\ell}(p_2)$};
    \vertex [below right=of z] (f8) at (16, -1.0) {\Large ${\ell}(p_3)$};
    \diagram* {
      (o)--[scalar, edge label={\Large $P$}](x),
      (x)--[photon, edge label={\Large $\gamma^*$}](y),
      (x)--[photon, edge label={\Large $\gamma^*$}](z),
      (f6)--[fermion](y)--[fermion](f5),
      (f7)--[fermion](z)--[fermion](f8),
    };
  \end{feynman}
\end{tikzpicture}
\end{minipage}
\caption{Feynman diagrams for $\mathcal{P} \to \ell^+\ell^-\ell^+\ell^-$.}
\label{fig_feyman}
\end{figure}

The coupling of the $\eta$ meson to virtual photons is an important input for calculating the anomalous magnetic moment of the muon, $a_{\mu} = (g_{\mu}-2)/2$, as pseudoscalar-meson exchange constitutes 
the dominant contribution to hadronic light-by-light scattering~\cite{muon}.
The double Dalitz decays $\mathcal{P} \rightarrow \ell^+\ell^-\ell^+\ell^-$, involving pseudoscalar mesons $\mathcal{P}$ ($\mathcal{P} = \pi^0, \eta$ or $\eta'$) and leptons $\ell$ ($\ell = e, \mu$), are expected to proceed through
two virtual photons~\cite{BESIII:2024ddb}. 
In such decays, each virtual photon internally converts into a lepton pair, as depicted in Fig.~\ref{fig_feyman}. In addition, the deviations between the measured rate and the pointlike quantum electrodynamics (QED) prediction are usually described in terms of a timelike transition form factor (TFF), which provides important information on the internal structure of the meson~\cite{Landsberg:1985gaz}. 
   
In 1967, initial QED predictions~\cite{Jarlskog:1967fpu} for double Dalitz decays of the $\eta$ meson were formulated, under the assumption that the TFF is unity. Since then, various theoretical models, 
including the hidden gauge model~\cite{HG1,HG2,HG3}, the modified vector meson dominance model (VMD)~\cite{m_VMD1,m_VMD2,m_VMD3,VMD} and a data driven approach~\cite{DDA}, have been applied to study the double Dalitz decays $\eta, \eta^\prime\rightarrow e^+e^-e^+e^-$. Despite these theoretical
advances, experimental results remain sparse. 
The first observation of $\eta\rightarrow e^+e^-e^+e^-$ was reported by the KLOE collaboration in 2011~\cite{KLOE:2011qwm}. More recently, BESIII observed the decay $\eta^\prime\rightarrow e^+e^-e^+e^-$~\cite{BESIII:2022cul}.  Table~\ref{list0} presents a comparison of predictions and measurements for the branching fractions of $\eta$ and $\eta^\prime\rightarrow e^+e^-e^+e^-$. Notably, the central value of the BESIII measurement for the $\eta^\prime\rightarrow e^+e^-e^+e^-$ decay is approximately twice that of the theoretical predictions, whereas no such discrepancy is observed for the $\eta$ decay.  
Compared to the KLOE measurement, the BESIII experiment benefits from larger $\eta$ samples 
produced in $J/\psi$ radiative and hadronic decays, providing unique access to $\eta$ decay channels, as demonstrated in Refs.~\cite{Fang:2021hyq,Kang:2023yye}. A new and high-precision measurement of these decays is therefore desirable. Such a measurement would not only validate the previous KLOE measurement, but also offer more experimental data to further improve our understanding of the relevant physical models. 

\begin{table}[htbp]
\begin{footnotesize}
\setlength{\tabcolsep}{4.5pt}
\begin{center}
\caption{The theoretical predictions and experimental measurements for the branching fractions of the decays $\eta\rightarrow e^+e^-e^+e^-$ and $\eta'\rightarrow e^+e^-e^+e^-$. The theoretical uncertainties are from input parameters, while the experimental uncertainties include both statistical and systematic components.}
\label{list0}
\begin{tabular}{ c c c}
     \hline
     \hline
     \multirow{2}{*}{Source}&  $\eta\to e^{+}e^{-}e^{+}e^{-}$ &$\eta' \to e^{+}e^{-}e^{+}e^{-}$\\
     & $(\times 10^{-5})$    &   $(\times 10^{-6})$ \\
     \hline
Hidden gauge~\cite{VMD}&  {$2.68\pm0.13$} &{$2.38\pm 0.40$}\\ 
Modified VMD~\cite{VMD}&  {$2.67\pm0.13$} &{$2.32\pm 0.40$}\\
Data driven approach~\cite{DDA}& {$2.17\pm0.20$} &{$2.10\pm 0.45$}\\
     \hline
KLOE~\cite{KLOE:2011qwm}&   {$2.40\pm0.22$}&$\cdots$\\
BESIII~\cite{BESIII:2022cul}& $\cdots$  & {$4.50\pm 0.12$}\\
    \hline
    \hline
\end{tabular}
\end{center}
\end{footnotesize}
\end{table}

In this paper, we present a measurement of the branching fraction for the double Dalitz decay of $\eta \to e^+e^-e^+e^-$, based on the dataset corresponding to $(1.0087 \pm 0.0044) \times {10^{10}}$ $J/{\psi}$ events~\cite{EVENTS} collected with the BESIII detector~\cite{ref::BesIII} at the BEPCII collider~\cite{ref::collider}.

{
\section{BESIII DETECTOR AND MONTE CARLO SIMULATION}
\label{sec:BES}
The BESIII detector~\cite{ref::BesIII} records symmetric $e^+e^-$ collisions provided by the BEPCII accelerator~\cite{ref::collider} in the center-of-mass energy ($\sqrt{s}$) range from 1.84 to 4.95~GeV, with a peak luminosity of $1.1 \times 10^{33}\;\text{cm}^{-2}\text{s}^{-1}$ achieved at $\sqrt{s} = 3.773\;\text{GeV}$. The cylindrical core of the BESIII detector covers 93\% of the full solid angle and consists of a helium-based multilayer drift chamber~(MDC), a plastic scintillator time-of-flight system (TOF) and a CsI (Tl) electromagnetic calorimeter (EMC), which are all enclosed in a superconducting solenoidal magnet providing a 1.0~T magnetic field (0.9~T in 2012, for 11\% of the data). The solenoid is supported by an octagonal flux-return yoke with resistive plate counter muon identifier modules interleaved with steel. The momentum resolution of the charged particle at $1~{\rm GeV}/c$ is $0.5\%$, and the specific ionization energy loss d$E$/d$x$ resolution is $6\%$ for the electrons from the Bhabha scattering. The EMC measures photon energies with a resolution of $2.5\%$ ($5\%$) at $1$~GeV in the barrel (end cap) region. The time resolution of the TOF barrel part is 68~ps, while that of the end cap part is 110~ps. The end cap TOF system was upgraded in 2015 using multigap resistive plate chamber technology, providing a time resolution of 60~ps, which benefits 85\% of the data used in this analysis~\cite{Tof1,Tof2,Tof3}.
\setlength{\parskip}{0.18em}

Simulated samples are produced with a {\sc geant4}-based~\cite{Geant4} Monte Carlo (MC) package. This package includes a geometric description of the BESIII detector and the detector response. These samples are used to determine the detection efficiency and estimate the backgrounds. The simulation includes the beam energy spread and initial state radiation (ISR) in the $e^+e^-$ annihilations modeled with the generator {\sc kkmc}~\cite{Jadach01}, while the decays are simulated using {\sc evtgen}~\cite{EVT}. 
Possible hadronic backgrounds are investigated using a sample of inclusive MC events of $J/{\psi}$ decay, where the known decays of $J/{\psi}$ are modeled with {\sc evtgen}~\cite{EVT} with branching fractions
taken from the Particle Data Group (PDG)~\cite{PDG}, while the remaining unknown decays are generated with the {\sc{lundcharm}} model~\cite{LUN}. 
To determine the detection efficiencies, we generate 2$\times$10$^6$ simulated signal events for each of the decay chains: $J/\psi \to \gamma \eta$ with $\eta \to e^+e^-e^+e^-$, and $J/\psi \to \gamma \eta'$ with $\eta' \to \pi^+\pi^-\eta$ followed by $\eta \to e^+e^-e^+e^-$. The signal generation employs an amplitude model based on the VMD framework~\cite{VMD}. The relevant background channels, $\eta \to \pi^+\pi^-e^+e^-$~\cite{DIY1} and $\eta \to \gamma e^+e^-$~\cite{DIY2}, are also included in the simulation. 
The dataset collected at $\sqrt{s} =$ 3.08 GeV with an integrated luminosity of (167.06 $\pm$ 0.10) pb$^{-1}$~\cite{EVENTS} is used to estimate the contamination from continuum processes.

\section{EVENTS SELECTION AND BACKGROUND ANALYSIS}
The $\eta$ candidates are reconstructed through the decays of $J/\psi \to \gamma \eta$~(mode I) and $J/\psi \to \gamma \eta'$, $\eta' \to \pi^+\pi^-\eta$ (mode II).
Charged tracks detected in the MDC must be within a polar angle range of $|{\rm cos}\,\theta|\leq 0.93$, where $\theta$ is defined with respect to the MDC axis. 
The distance of closest approach to the interaction point (IP) must be less than 10 cm along the $z$ direction and less than 1 cm in the plane transverse to the $z$-axis.
Particle identification (PID) for charged tracks combines with the measurements of the specific ionization energy loss in the MDC (d$E$/d$x$) and the flight time in the TOF to form likelihoods of the electron hypotheses, $\mathcal{L}(e)$, and pion, $\mathcal{L}(\pi)$, hypotheses. Tracks are identified as electrons when $\mathcal{L}(e) > \mathcal{L}(\pi)$, while charged pions are identified by requiring $\mathcal{L}(\pi)>\mathcal{L}(e)$. To study $\eta \to e^{+}e^{-}e^{+}e^{-}$ via mode~I and mode~II, the $e^{+}e^{-}e^{+}e^{-}$ and $\pi^{+}\pi^{-}e^{+}e^{-}e^{+}e^{-}$ combinations are selected for further analysis, respectively.

Photons are reconstructed from isolated showers in the electromagnetic calorimeter (EMC). The energy deposited in the nearby time of flight counter is incorporated to enhance the reconstruction efficiency and energy resolution. The energies of photons are required to be greater than 25 MeV in the EMC barrel region ($|{\rm cos}\,\theta|<0.80$), and greater than 50 MeV in the EMC end-cap region ($0.86<|{\rm cos}\,\theta|<0.92$). Moreover, to suppress electronic noise and showers not related to the event, the difference between the EMC time and the event start time must fall within the range (0, 700) ns. To exclude showers that originate from charged tracks, the angle between the direction of each shower in the EMC and the direction of the closest extrapolated charged track must be greater than 15 degrees. Events with at least one photon are retained for further analysis.

To further suppress potential backgrounds and improve mass resolution, a four-constraint (4C)  kinematic fit is implemented for all the $\gamma e^+e^-e^+e^-$ combinations for the process $J/\psi \to \gamma \eta$, enforcing energy-momentum conservation from the initial $e^+e^-$ state. Candidates with the minimum $\chi^2_{\rm 4C+PID}(e^+e^-e^+e^-)$ per event are retained, where $\chi^2_{\rm 4C+PID}=\chi^2_{4\rm{C}}+\Sigma^4_{i=1}\chi^2_{\rm{PID}}(i)$, is the sum of $\chi^2$ calculated from the 4C kinematic fit ($\chi^2_{4\rm{C}}$) and the PID ($\chi^2_{\rm{PID}}$). 
For the $J/\psi \to \gamma \eta'$, $\eta' \to \pi^+\pi^-\eta$, $\eta \to e^+e^-e^+e^-$ channel, a five-constraint (5C) kinematic fit is implemented for all the $\gamma\pi^+\pi^- e^+e^-e^+e^-$ combinations, with an additional mass constraint on the 
$\pi^+\pi^-\eta$ system based on the known $\eta'$ mass~\cite{PDG}.
Candidates with the minimum $\chi^2_{\rm 5C}(\pi^+\pi^-e^+e^-e^+e^-)$ are retained.

The requirements of $\chi^2_{\rm 4C+PID}(e^+e^-e^+e^-)<60$ and $\chi^2_{\rm 5C}(\pi^+\pi^-e^+e^-e^+e^-)<60$ were obtained by optimizing the figure-of-merit $S/\sqrt{S+B}$, where $S$ ($B$) represents the number of signal 
(background) events from MC simulations. 
The dominant backgrounds are found to originate from $\eta\to\gamma e^+e^-$ and $\eta\to\gamma\gamma$ decays, where a photon converts to an $e^+e^-$ pair at the beam pipe or the inner wall of the MDC. 
Fig.~\ref{fig_beforecutgconv}(a) shows the invariant mass distribution of the $e^+e^-$ pairs, with a clear peak around 
0.015~GeV/$c^2$ corresponding to the $\gamma$ conversion background.

\begin{figure}[h]
\begin{center}
\begin{minipage}[t]{1.0\linewidth}
\includegraphics[width=1\textwidth]{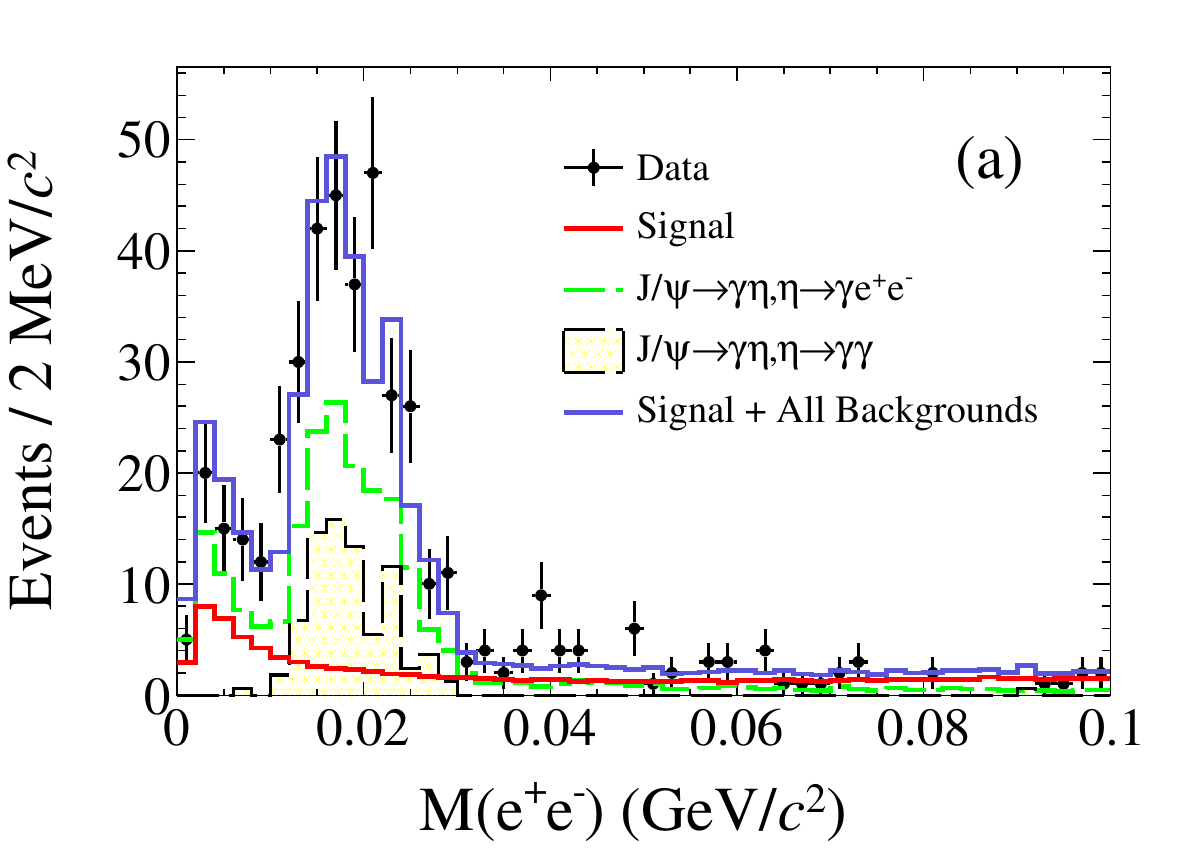}
\end{minipage}
\begin{minipage}[t]{1.0\linewidth}
\includegraphics[width=1\textwidth]{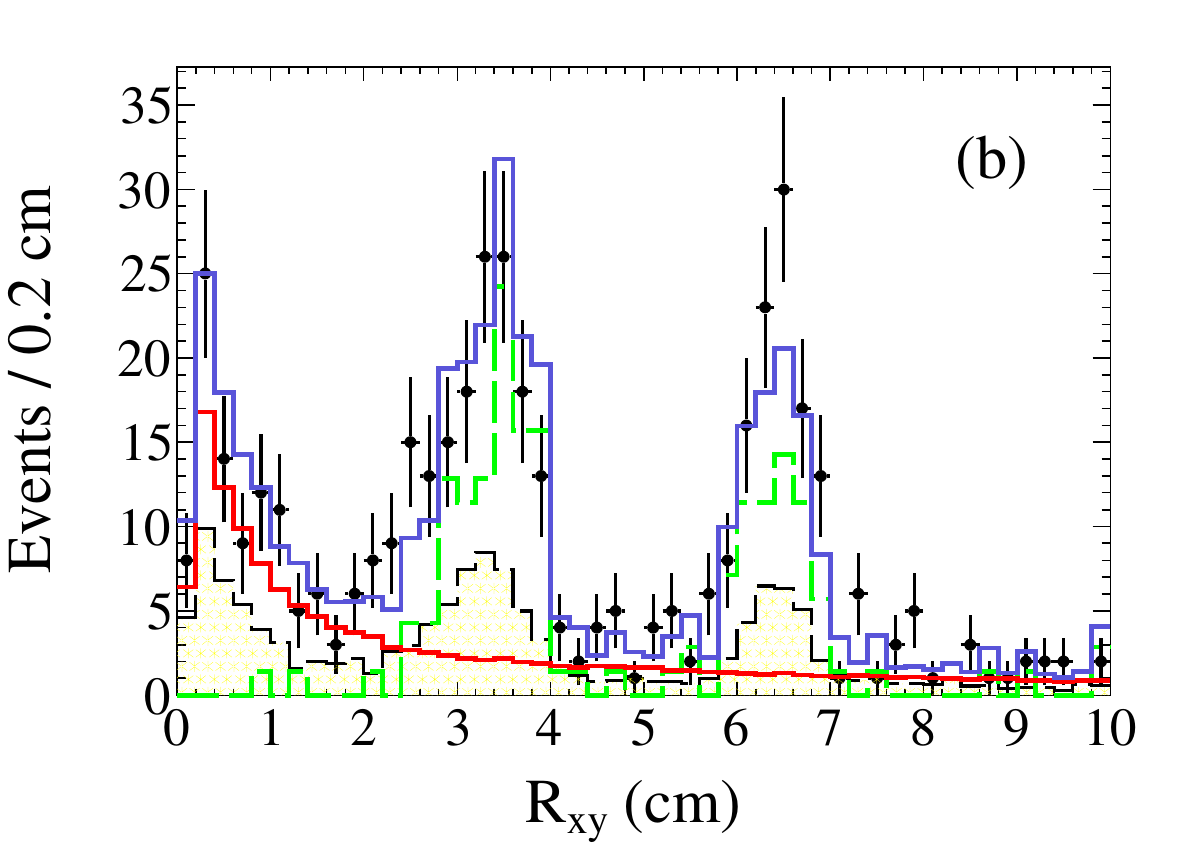}
\end{minipage}
\caption{The distributions of (a) $M_{ee}$ and (b) $R_{xy}$ for mode I. The dots with error bars are data, the red histograms are the signal MC events, the shaded yellow histograms are the background of $J/\psi \to \gamma \eta, \eta\to\gamma e^{+}e^{-}$ and the dashed green histograms are the background of $J/\psi \to \gamma \eta, \eta\to \gamma\gamma$. The blue histograms are the sum of all backgrounds normalized from the MC simulation. }
\label{fig_beforecutgconv}
\end{center}
\end{figure}

To distinguish the $\gamma$ conversion events from signal events, the photon conversion finder package~\cite{Xu:2012xq} is used. 
Two variables are reconstructed: (i) the distance $R_{xy}$ from the $e^+e^-$ vertex (of the $\theta_{ee}^1$ pair; see below) to the IP in the transverse plane, as shown in Fig.~\ref{fig_beforecutgconv}(b);
(ii) the opening angle $\Phi_{ee}$ of the $e^+e^-$ pair in the laboratory frame~\cite{PHENIX:2009gyd}, as shown in Figs.~\ref{fig_cut}(a),~\ref{fig_cut}(b),~\ref{fig_cut}(c) and~\ref{fig_cut}(d). 
As shown in Fig.~\ref{fig_beforecutgconv}(b), the signal $e^+e^-$ pairs in the $R_{xy}$ distribution originate from the IP, 
while the $\gamma$ conversion events peak around 3~cm in the beam pipe and 6~cm in the inner wall of the MDC.
For the $\gamma$ conversion events, $\Phi_{ee}$ is expected to be close to zero.
We therefore veto the $\gamma$ conversion background by requiring $\Phi_{ee} < 70^{\circ}$ and 2 cm $< R_{xy} <$ 8 cm for mode I, and $\Phi_{ee} < 40^{\circ}$ and 2 cm $< R_{xy} <$ 8 cm for mode II, as illustrated in Figs.~\ref{fig_cut}(a),~\ref{fig_cut}(b),~\ref{fig_cut}(c) and~\ref{fig_cut}(d).

\begin{figure*}[hptb]
\begin{center}
\begin{minipage}[t]{0.45\linewidth}
\includegraphics[width=1\textwidth]{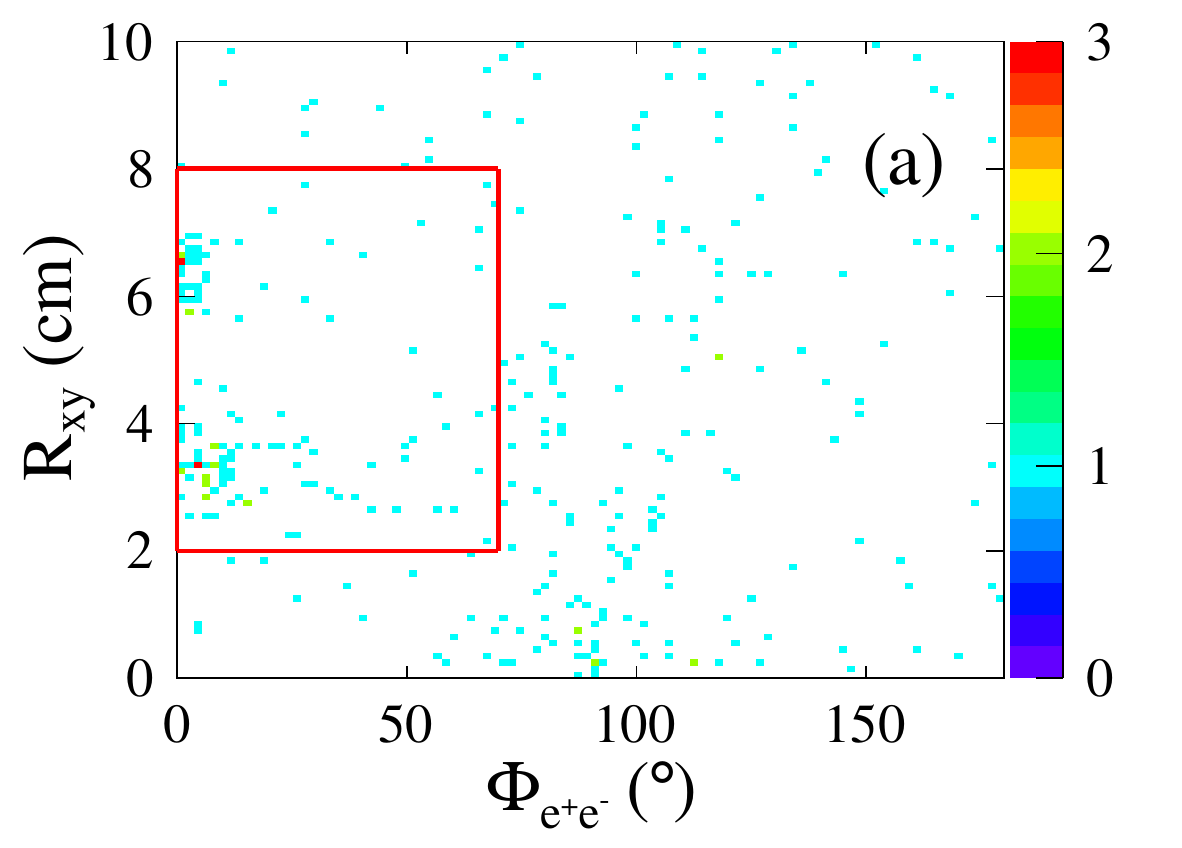}
\end{minipage}
\begin{minipage}[t]{0.45\linewidth}
\includegraphics[width=1\textwidth]{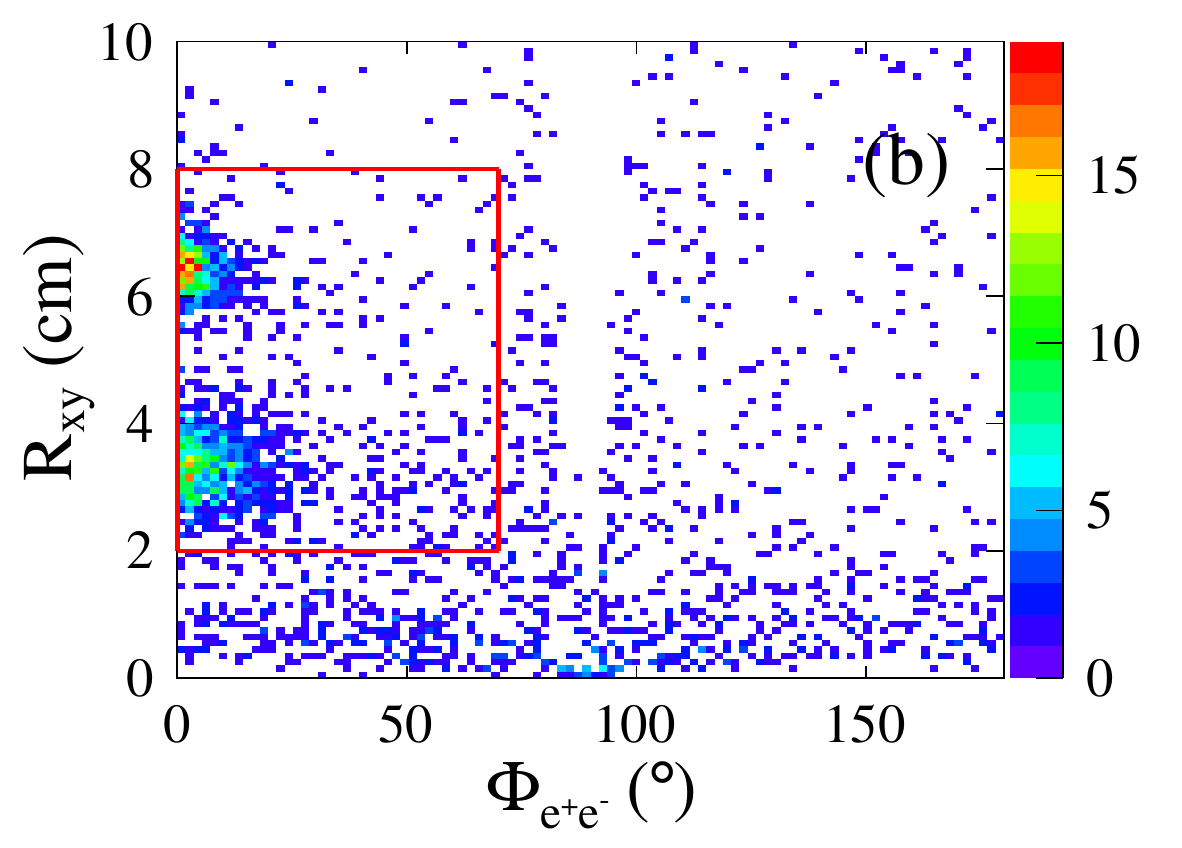}
\end{minipage}
\begin{minipage}[t]{0.45\linewidth}
\includegraphics[width=1\textwidth]{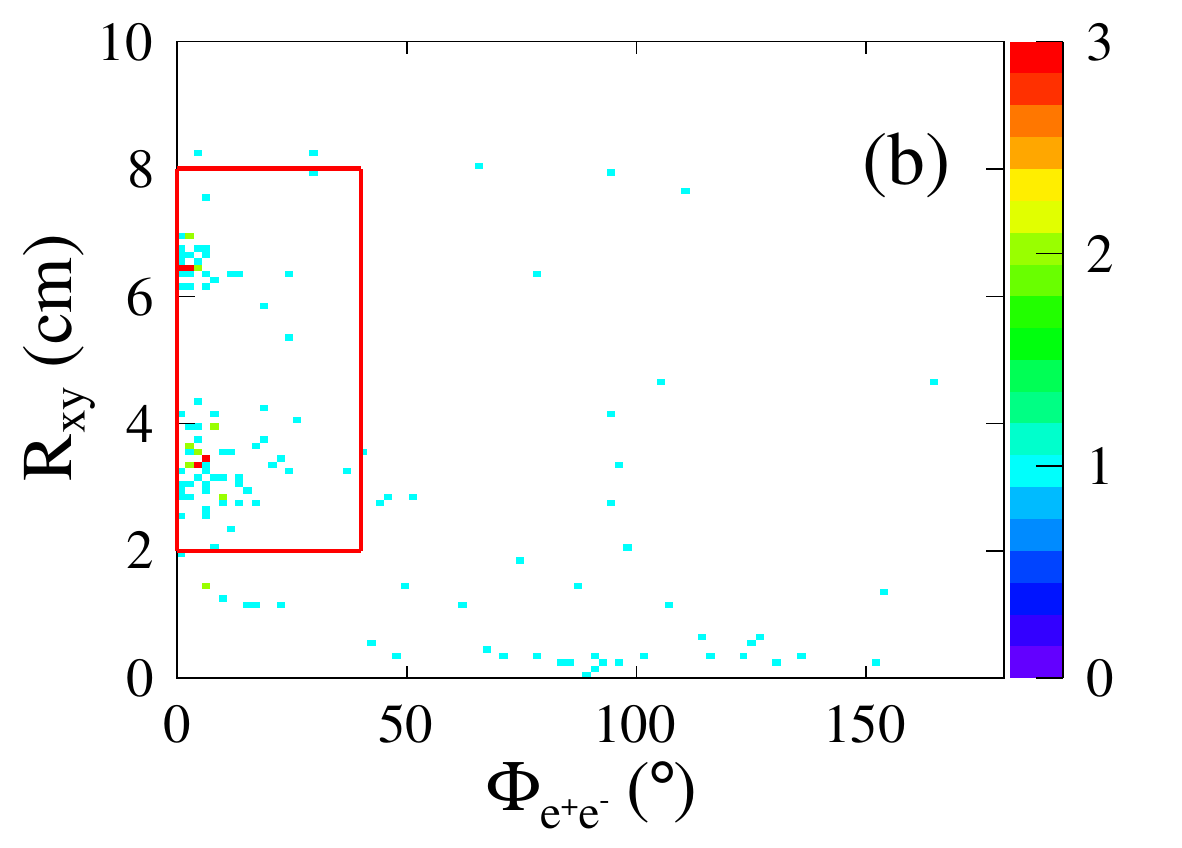}
\end{minipage}
\begin{minipage}[t]{0.45\linewidth}
\includegraphics[width=1\textwidth]{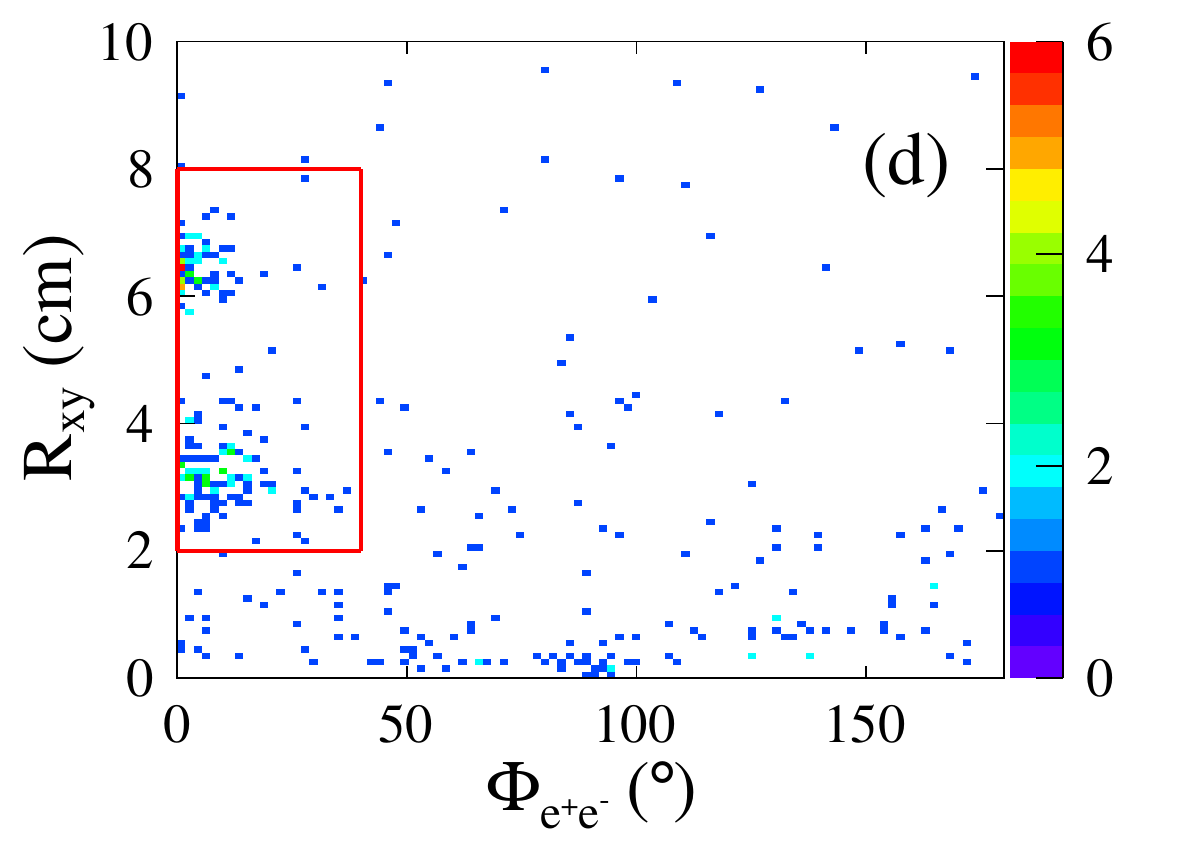}
\end{minipage}
\begin{minipage}[t]{0.45\linewidth}
\includegraphics[width=1\textwidth]{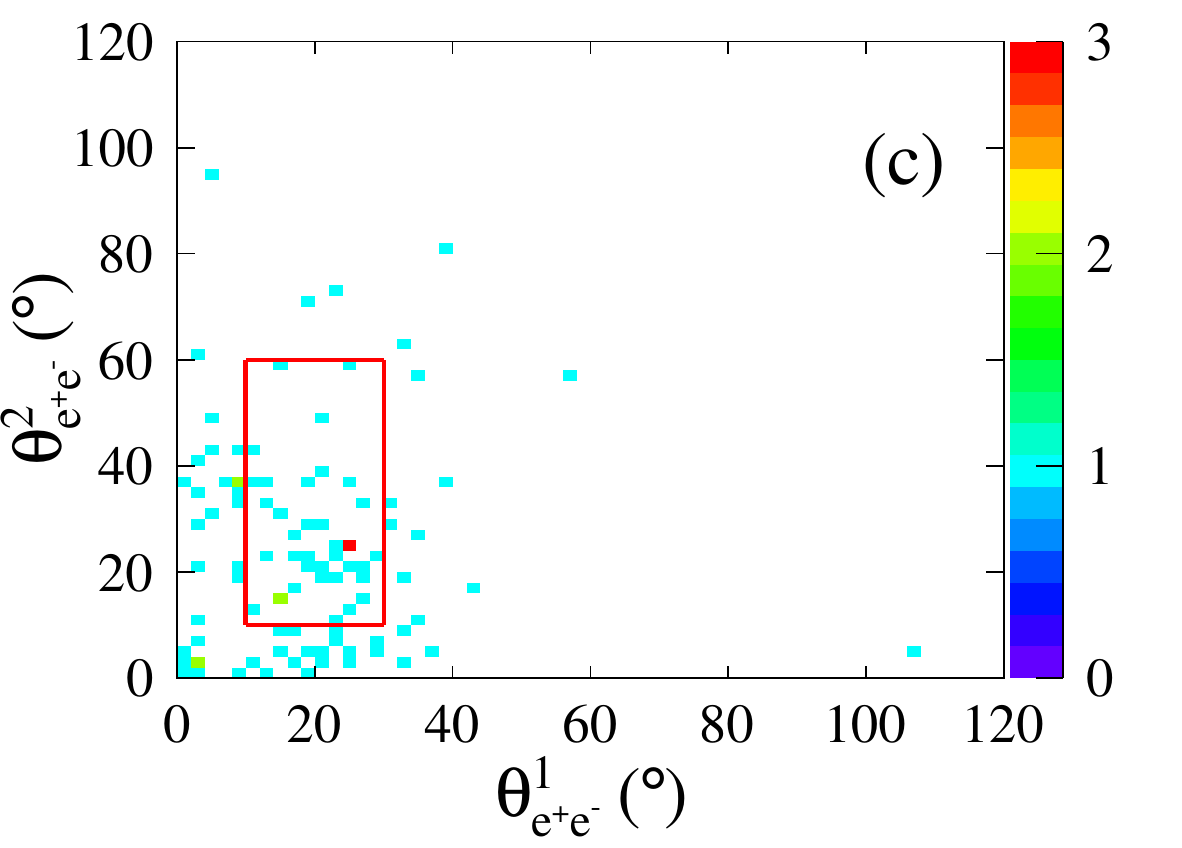}
\end{minipage}
\begin{minipage}[t]{0.45\linewidth}
\includegraphics[width=1\textwidth]{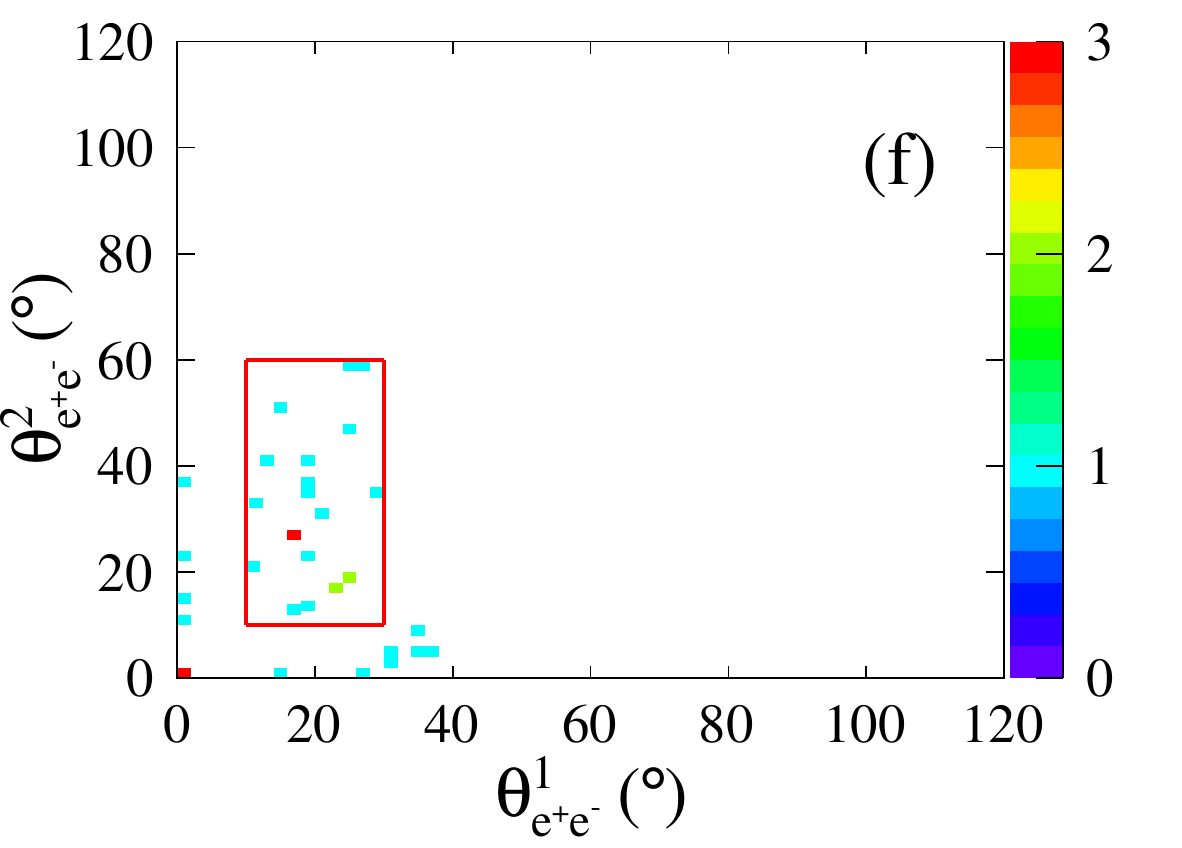}
\end{minipage}
\begin{minipage}[t]{0.45\linewidth}
\includegraphics[width=1\textwidth]{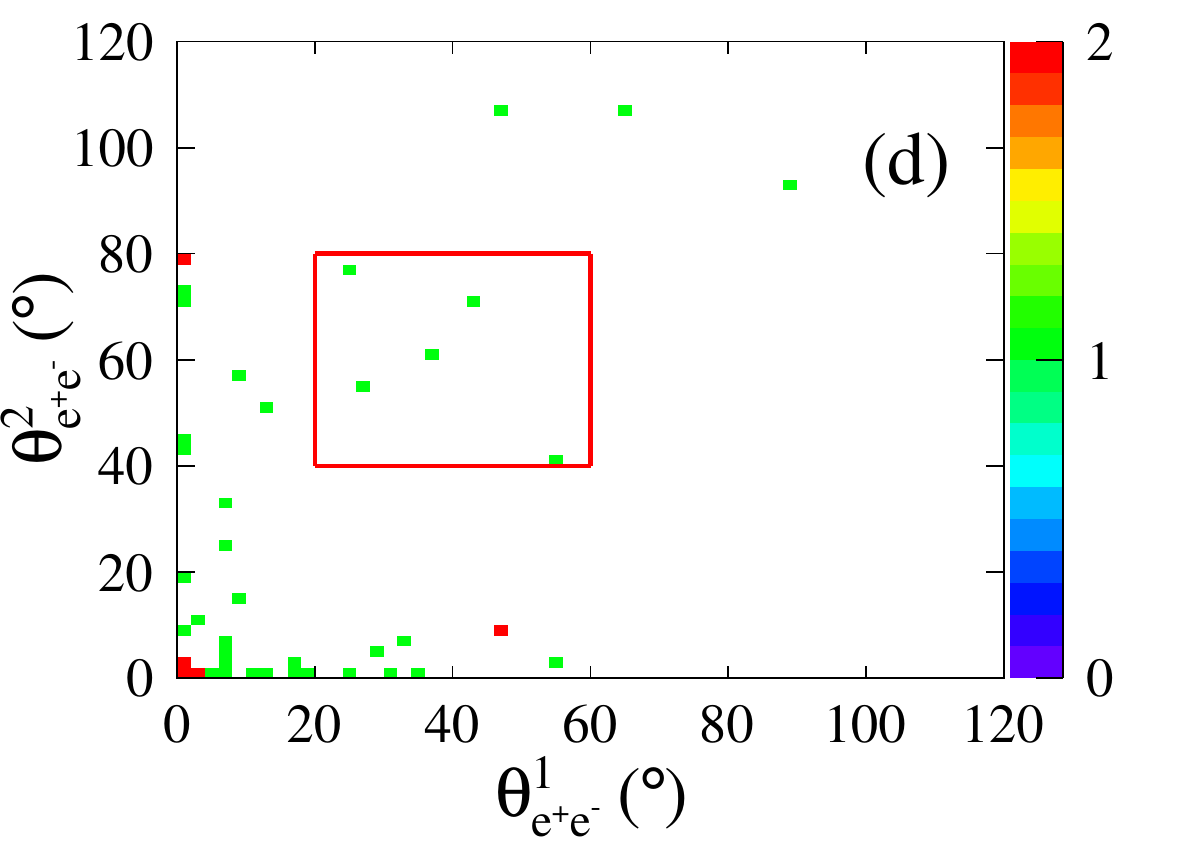}
\end{minipage}
\begin{minipage}[t]{0.45\linewidth}
\includegraphics[width=1\textwidth]{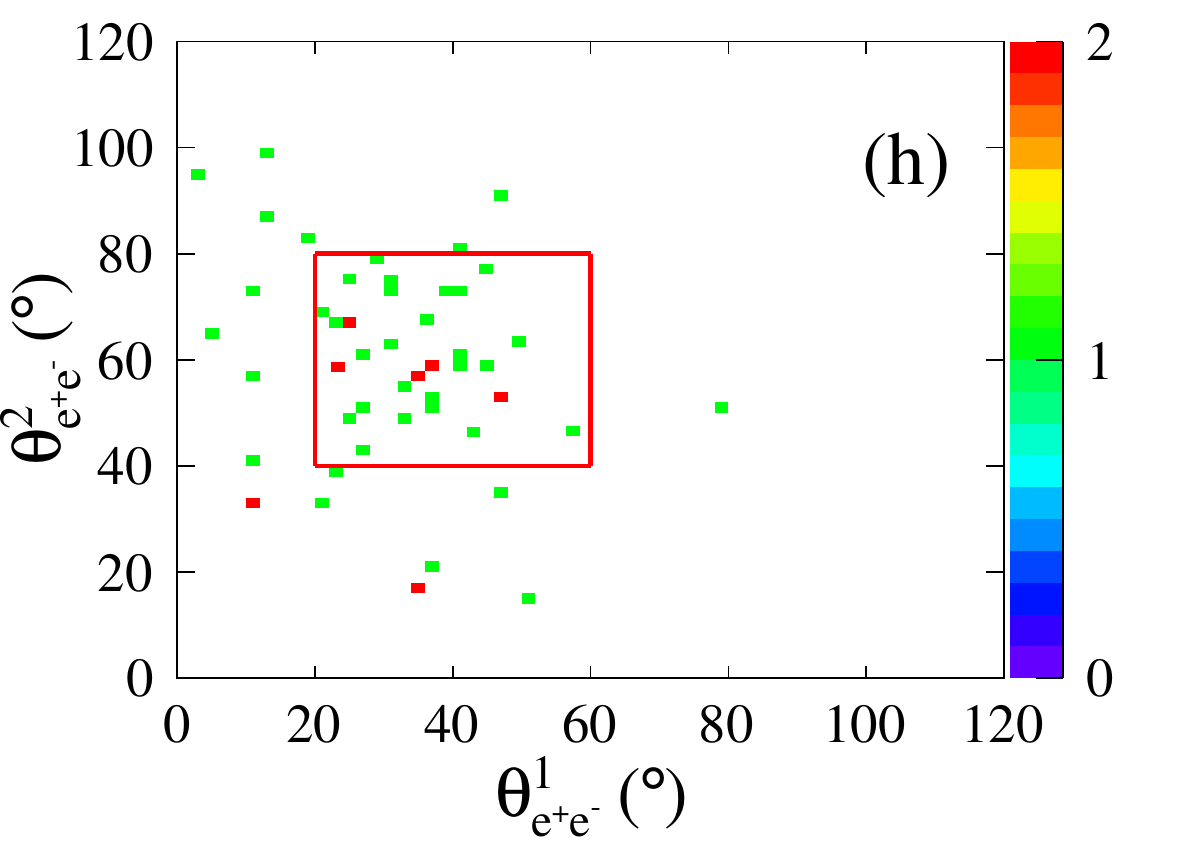}
\end{minipage}
\caption{The photon conversion veto criterion in the $\Phi_{ee}$ vs. $R_{xy}$ plane, shown with events from (a)(c) data and (b)(d) $\eta\to\gamma e^+e^-$ MC for mode I and mode II, respectively. The miscombination background veto criterion in the $\theta_{ee}^{1}$ vs. $\theta_{ee}^{2}$ plane, shown with events from (e)(g) data and (f)(h) $\eta\to\pi^+\pi^-e^+e^-$ MC for mode I and mode II, respectively. The red boxes denote the veto regions.}
\label{fig_cut}
\end{center}
\end{figure*}

Another significant background arises from the misidentification background of $\pi$ as $e$, such as $\eta\to \pi^{+}\pi^{-}e^{+}e^{-}$.
To suppress this background, we use the opening angles $\theta_{ee}$ between the $e^+e^-$ pairs in the laboratory frame, 
defined as $\cos \theta_{ee} = \hat{p}_+ \cdot \hat{p}_-$, where $\hat{p}_+$ and $\hat{p}_-$ are the normalized three-momenta (directions) of one of the positrons and one of the electrons, respectively.
For signal events, $\theta_{ee}$ is expected to be close to zero,
for correctly paired (i.e., from the same photon) tracks.  
To veto the particle misidentification background, two angles between the two electrons are defined: the smallest angle of the four $e^+e^-$ pairings, $\theta_{ee}^{1}$, and the angle of the complementary pair, $\theta_{ee}^{2}$. 
We reject events with $10^\circ < \theta_{ee}^{1} < 30^\circ$ when $10^\circ < \theta_{ee}^{2} < 60^\circ$ for mode I, and $20^{\circ}< \theta_{ee}^{1} < 60^{\circ}$ when $40^{\circ}< \theta_{ee}^{2} < 80^{\circ}$ for mode II, as shown in Figs.~\ref{fig_cut}(e),~\ref{fig_cut}(f),~\ref{fig_cut}(g) and~\ref{fig_cut}(h).

Other potential backgrounds from $J/\psi$ decays and continuum production are studied with the $J/\psi$ inclusive MC sample and the data samples taken at $\sqrt{s}=3.08$~GeV, respectively. The main background channels are summarized in Table~\ref{bkgs}. After performing a study of the exclusive MC simulations for these survived backgrounds, their contributions are normalized according to the branching fractions taken from the PDG~\cite{PDG}, and fixed in the subsequent fit. The continuum background is found to be negligible.

\begin{table}[!htp]
\begin{footnotesize}
\setlength{\tabcolsep}{0.5pt}
\caption{The dominant background channels and the normalized yields for mode I and mode II. A dash indicates that the corresponding contribution is free in the fit. The uncertainties are statistical.}
    \centering
    \begin{tabular}{lcc}
    \hline \hline
    Mode &Background channel &Normalized yield \\
    \hline
    \multirow{3}{*}{I}     &$J/\psi \to \gamma \eta$, $\eta \to \gamma e^+ e^-$ &$4.8 \pm 0.3$   \\
         &$J/\psi \to \gamma \eta$, $\eta \to \pi^+ \pi^- e^+ e^-$ &$2.8 \pm 0.2$ \\
         &$J/\psi \to \gamma \eta$, $\eta \to \gamma \gamma$ & $\cdots$ \\
         \hline
    \multirow{3}{*}{II} &$J/\psi \to \gamma \eta'$, $\eta' \to \pi^+ \pi^- \eta$, $\eta \to \gamma e^+ e^-$ &$8.6 \pm 0.6$      \\
    II  &$J/\psi \to \gamma \eta'$, $\eta' \to \pi^- \pi^+ \eta$, $\eta \to \pi^+ \pi^- e^+ e^-$ &$1.6 \pm 0.1$ \\
    &$J/\psi \to \gamma \eta'$, $\eta' \to \pi^+ \pi^- \eta$, $\eta \to \gamma \gamma$ & $\cdots$ \\
    \hline \hline
    \end{tabular}
    \label{bkgs}
    \end{footnotesize}
\end{table}

\section{MEASUREMENT OF BRANCHING FRACTION}
The signal yields are extracted through a simultaneous unbinned extended maximum-likelihood fit to the $M(e^+e^-e^+e^-)$ distributions of both decay modes, with a shared branching fraction for $\eta \to e^+e^-e^+e^-$. Due to limited data statistics, the signal probability density functions (PDFs) are derived from MC simulation. The backgrounds from the decays of $\eta \to \pi^{+}\pi^{-}e^{+}e^{-}$ and $\eta \to \gamma e^{+}e^{-}$ in both modes are modeled with the MC-simulated shapes, with yields fixed according to known branching fractions~\cite{PDG}. The remaining background (RBKG), related to the $e$, $\mu$ or $\pi$ misidentification and the QED processes, is found to distribute smoothly in the $M(e^+e^-e^+e^-)$ distribution and therefore modeled by a first-order polynomial function. Figure.~\ref{fig4e:fit} shows the simultaneous fit to the $M(e^+e^-e^+e^-)$ distributions for both modes. 
\begin{figure*}[hptb]
\begin{center}
\begin{minipage}[t]{0.48\linewidth}
\includegraphics[width=1\textwidth]{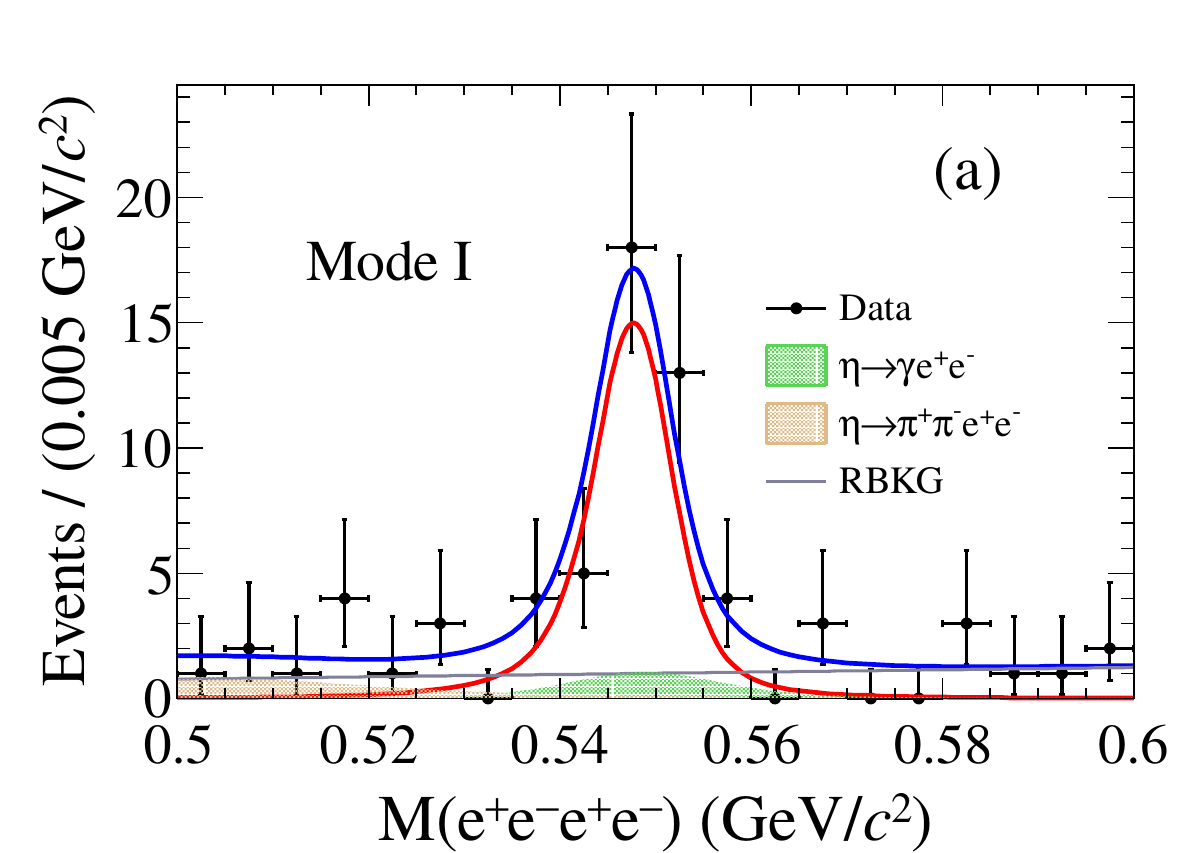}
\end{minipage}
\begin{minipage}[t]{0.48\linewidth}
\includegraphics[width=1\textwidth]{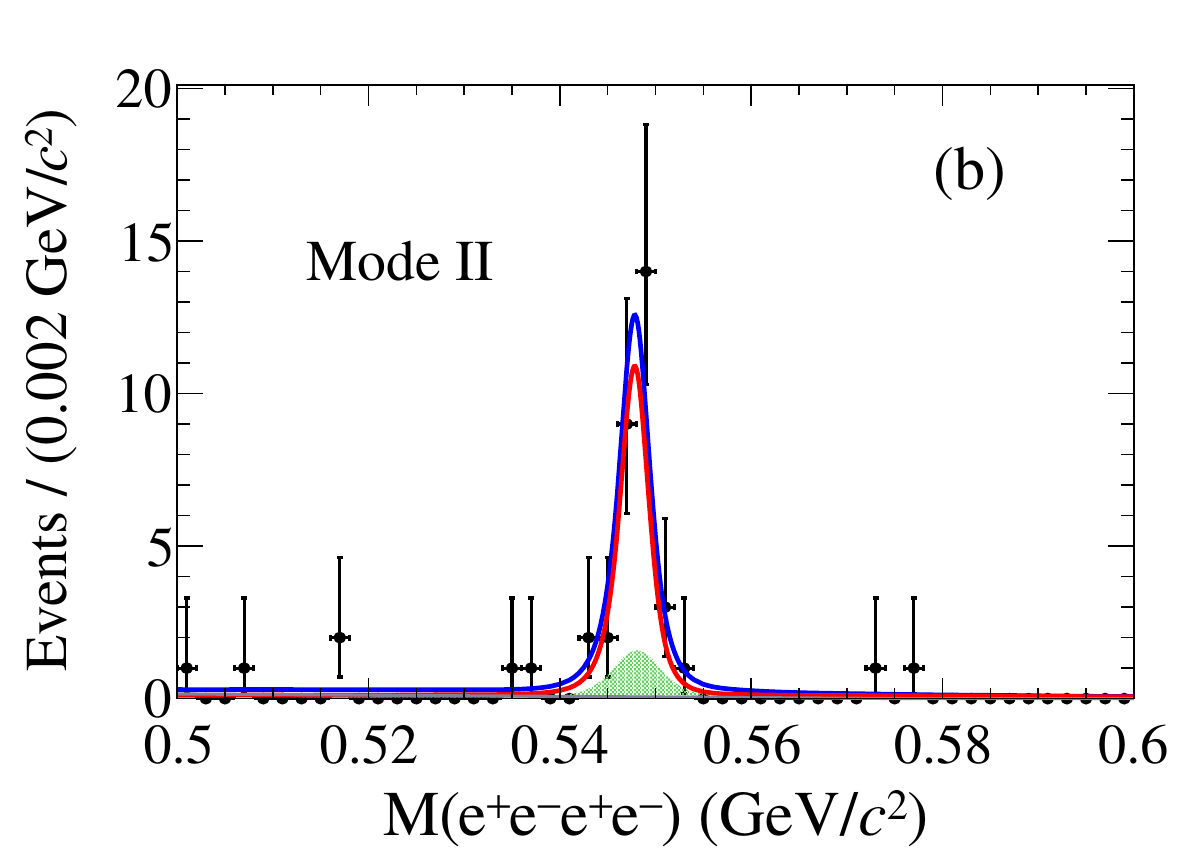}
\end{minipage}
\caption{Fits to the distributions of $M(e^{+}e^{-}e^{+}e^{-})$ for (a) mode I and (b) mode II. The dots with error bars are data, 
the blue curves are the fit results, the dashed red lines are the signal shapes, the hatched histograms are the backgrounds from $\eta \to \pi^{+}\pi^{-}e^{+}e^{-}$ and $\eta \to \gamma e^{+}e^{-}$ for both modes, and the gray lines are the smooth backgrounds.}
\label{fig4e:fit}
\end{center}
\end{figure*}
The statistical significances for $\eta\to e^+e^-e^+e^-$ are calculated to be $5.9\sigma$ for mode I and $7.8\sigma$ for mode II from the independent fits, respectively, and $11.1\sigma$ from the simultaneous one. They are calculated from the changes in log-likelihood and degrees of freedom with and without the signal in the fit. 

The branching fraction of $\eta\to e^{+}e^{-}e^{+}e^{-}$ is then determined by
\begin{linenomath*}
\begin{dmath}
\mathcal{B}({\eta}\to e^{+}e^{-}e^{+}e^{-})= {N^{\rm obs} \over {{N_{J/{\psi}}} \cdot \epsilon \cdot {\mathcal{B}}}},
\label{BR1}
\end{dmath}
\end{linenomath*}
where $N^{\rm obs}$ is the number of signal events in fit, $N_{J/\psi}$ is the total number of $J/\psi$ events in data, and $\epsilon$ denotes the detection efficiency. The $\mathcal{B}$ denotes the branching fraction of $J/\psi\to\gamma \eta$ for mode I and the product of the branching fractions of $J/\psi\to\gamma \eta'$ and $\eta'\to\pi^+\pi^-\eta$ for mode II; these values are taken from the PDG~\cite{PDG}. Table~\ref{weight} summarizes the numerical results. From the simultaneous fit to both modes, the branching fraction of $\eta\to e^{+}e^{-}e^{+}e^{-}$ is determined to be $(2.63 \pm 0.34)\times {10^{-5}}$, where the uncertainty is statistical only. 
\begin{table}[htbp]
  \caption{Numerical results for signal yields, $N^{\rm obs}$, detection efficiencies, $\epsilon$, branching fractions of $\eta \to e^+e^-e^+e^-$, $\mathcal{B}$, and statistical significances, $S$, for both modes.}
\setlength{\tabcolsep}{9pt} 
  \begin{tabular} {c c c c c c} \hline\hline
Mode &$N^{\rm obs}$   & $\epsilon$~(\%) &$\mathcal{B}~(\times 10^{-5})$ & $S$~($\sigma$) \\ \hline

I  & $35.9\pm 6.7$   &  13.1     &$2.51 \pm 0.47$   & 5.9\\
II &$29.5\pm 6.1$  &   4.7     &$2.80 \pm 0.58$  &7.8\\

   \hline
   \hline
  \end{tabular}
  \label{weight}
\end{table}

\section{SYSTEMATIC UNCERTAINTY}
\label{sec:systematics}
Systematic uncertainties on the branching fraction measurements mainly originate from
tracking and PID, 
photon detection, 
kinematic fit, 
$\gamma$ conversion veto, 
MC generator,
signal shape,
background shape, 
fit range,
 branching fractions of intermediate decays, total number of $J/{\psi}$ events.

\begin{enumerate}[label=(\roman*)]
\item 
The systematic uncertainties associated with the tracking and PID for pions are studied using a control sample of $J/\psi \to \pi^+ \pi^- \pi^0$, and those for $e^\pm$ are evaluated using a mixed sample of $J/\psi\to e^+ e^- \gamma$ and $J/\psi \to e^+ e^- \gamma_{\rm FSR}$ (FSR denotes final state radiation). In both cases, the difference in efficiencies between data and MC simulation is extracted as a function of the particle momentum and the cosine of its polar angle. Subsequently, each event in the signal MC samples is reweighted by a factor $(1+\Delta_{\rm syst})$. Then the branching fractions are recalculated with the detection efficiencies determined from the reweighted signal MC sample. The resulting differences compared to the original ones are taken as individual systematic uncertainties. 
\item 
The systematic uncertainty of photon detection is studied using a control sample of $e^+e^-\to\gamma_{\rm{ISR}}\mu^+\mu^-$. The difference in the detection efficiencies between data and the MC simulation is assigned as the systematic uncertainty. 
\item 
The systematic uncertainty associated with the kinematic fit is assigned by comparing the results with and without track helix parameter corrections~\cite{4C}. Half of the difference in the detection efficiency is taken as the systematic uncertainty.
\item 
The systematic uncertainty from the $\gamma$ conversion veto is studied with a control sample of $J/\psi\to\pi^+\pi^-\pi^0$, $\pi^0\to\gamma e^+e^-$~\cite{DIY2}. The differences of the detection efficiencies between data and MC simulation associated with the photon conversion veto are assigned as the systematic uncertainties.
\item 
The MC generator, based on the theoretical calculation in Ref.~\cite{DIY1}, is used to determine the detection efficiency. The dependence of detection efficiency on the form factor is evaluated by varying the $c_3$ parameter in the modified VMD amplitude factor from 1. in the hidden-gauge model to 0.927-0.930 in the modified VMD model~\cite{VMD}
\begin{linenomath*}
\begin{dmath}
        \text{VMD}(s_{12},s_{34},s_{23}, s_{14}) \\
        = 1 - c_3 + c_3 \,\, \frac{m^2_V}{m^2_V - s_{12} - im_V\Gamma(s_{12})} \\
        \frac{m^2_V}{m^2_V - s_{34} - im_V\Gamma(s_{34})} \,\, \frac{m^2_V}{m^2_V - s_{23} + im_V\Gamma(s_{23})}\\
        \frac{m^2_V}{m^2_V - s_{14} + im_V\Gamma(s_{14})}.
        \label{c3}
\end{dmath}
\end{linenomath*}
Here the $s_{ij}$ = $(P_{e^+}+P_{e^-})^2$ are squares of the pairwise invariant masses, $m_V$ is the vector meson mass, and $\Gamma(s_{ij})$ is its total width~\cite{Qin:2017vkw}
\begin{linenomath*}
\begin{align}
\Gamma(s_{ij}) &=  \frac{g \sqrt{s_{ij}}}{m_V}  \left[ \left(1 - \frac{4m_e^2}{s_{ij}} \right) / \left(1 - \frac{4m_e^2}{m_V^2} \right) \right]^{\frac{3}{2}} \notag \\
&\quad \, \Theta(s_{ij}-4m_e^2),
\end{align}
\end{linenomath*}
where $g = 149.1$ MeV and $\Theta$ is the Heaviside step function. The maximum difference of the detection efficiencies between the nominal and alternative models is taken as the uncertainty due to the MC generator.
\item 
The systematic uncertainty due to the signal shape is estimated by changing the nominal signal function to a double Gaussian function. The difference of the signal yield is taken as the systematic uncertainty. 
\item 
The systematic uncertainty due to the fit range of $M(e^{+}e^{-}e^{+}e^{-})$ is estimated by varying the mass range by 10 or 20 MeV/$c^{2}$. The largest resulting difference in the signal yield is taken as the systematic uncertainty.
\item 
In the fit, the numbers of events for the backgrounds from 
$\eta \to \pi^{+}\pi^{-}e^{+}e^{-}$ and $\eta \to \gamma e^{+}e^{-}$ 
are fixed according to the branching fractions from the PDG. 
The associated systematic uncertainty is evaluated by randomly varying those branching fractions based on their uncertainties and repeating the simultaneous fit.
The resulting spread of the signal yield is taken as the systematic uncertainty.
To evaluate the uncertainties of the used branching fractions, a set of random numbers are generated within the uncertainty of each branching fraction. 
The uncertainty due to the assumed polynomial background shape is estimated by an alternate fit using a second-order polynomial function. 
\item 
The systematic uncertainties associated with the branching fractions of intermediate decays $J/{\psi} \to \gamma \eta$, $J/{\psi}\to \gamma \eta'$ and $\eta' \to \pi^+\pi^-\eta$ are taken from the PDG~\cite{PDG}, respectively. 
\item 
The total number of $J/\psi$ events ($N_{J/\psi}$) is determined with inclusive hadronic $J/\psi$ decays, with an uncertainty of 0.4\% ~\cite{EVENTS}.
\end{enumerate}

Assuming all sources are independent, the total systematic uncertainty is calculated as the square root of their quadratic sum. 
The various systematic uncertainties on the branching fraction measurements are summarized in Table~\ref{list_sys}. 
Note that the combined systematic uncertainty for both modes is propagated by $\sqrt{\sum_{i}\sigma^{2}_{i}\cdot w^{2}_{i} + \delta^{2}}$. Here, $\sigma_{i}$ denotes the individual systematic uncertainty for each mode, and $w_{i}$ is the weight factor defined as a ratio of the number of signal events for an individual mode to the sum of those for both modes, i.e., $w_{i} = N_{i}^{\rm obs}/(N_{\rm I}^{\rm obs}+N_{\rm II}^{\rm obs})$. The $\delta$ is the uncertainty of the simultaneous fit, which includes the signal shape, the background shape, and the fit range.

\begin {table}[htbp]
\begin{center}
\caption {Systematic uncertainties on the measurement of the branching fraction, in \%.}
\label{list_sys}
\begin {tabular}{l c c c c}
\hline\hline
Source      &Mode I           &Mode II      &Combined \\ \hline
Tracking                                 &  3.2            &   5.6         &      4.2    \\
PID                                          &  1.7            &   1.6         &      1.7   \\
Photon detection                             &  0.5            &   0.5         &      0.5    \\
Kinematic fit                                &  1.0            &   1.7         &      1.3   \\
Photon conversion veto                       &  3.0            &   3.0         &      3.0    \\
MC Generator                              &  0.8            &   1.1         &      0.9    \\
Signal shape                                 &  1.9            &   0.4         &      0.8    \\
Fit range                                    &  3.6            &   3.2         &      1.9    \\
Background shape                             &  1.2            &   0.2         &      0.7    \\
$\mathcal {B}(J/{\psi}\to\gamma \eta)$	     &  1.7            &   $\cdots$    &      1.0  \\
$\mathcal {B}(J/{\psi}\to\gamma \eta')$	     &  $\cdots$       &  1.3          &      0.5  \\
$\mathcal {B}(\eta' \to \pi^+\pi^-\eta)$	 &  $\cdots$       &   1.2         &      0.5   \\
Number of $J/{\psi}$ events                  &  0.4            &  0.4          &      0.4    \\
\hline
Total                                        &  6.7            &   7.8         &      6.2         \\
\hline
\hline
\end{tabular}
\end{center}
\end{table}

\section{SUMMARY}
Based on $(1.0087 \pm 0.0044) \times 10^{10}$ $J/\psi$ events collected with the BESIII detector, we have measured the branching fraction of $\eta \to e^+e^-e^+e^-$ via the decay chains $J/\psi \to \gamma\eta$ and $J/\psi \to \gamma\eta'$, $\eta' \to \pi^+\pi^-\eta$. The observed signals have statistical significances of $5.9\sigma$ and $7.8\sigma$, respectively. The combined branching fraction is determined to be $(2.63 \pm 0.34 \pm 0.16) \times 10^{-5}$, where the first uncertainty is statistical and the second is systematic. As shown in Fig.~\ref{result}, this result is consistent with the previous KLOE measurement~\cite{KLOE:2011qwm} and theoretical predictions~\cite{VMD,DDA}. Future experiments at facilities like the super $\tau$-charm facility~\cite{Achasov:2023gey} and the JLab Eta Factory~\cite{Somov:2024jiy} will enable higher-precision studies of this decay.

\begin{figure}[hptb]
\begin{center}
\begin{minipage}[t]{1.0\linewidth}
\includegraphics[width=1\textwidth]{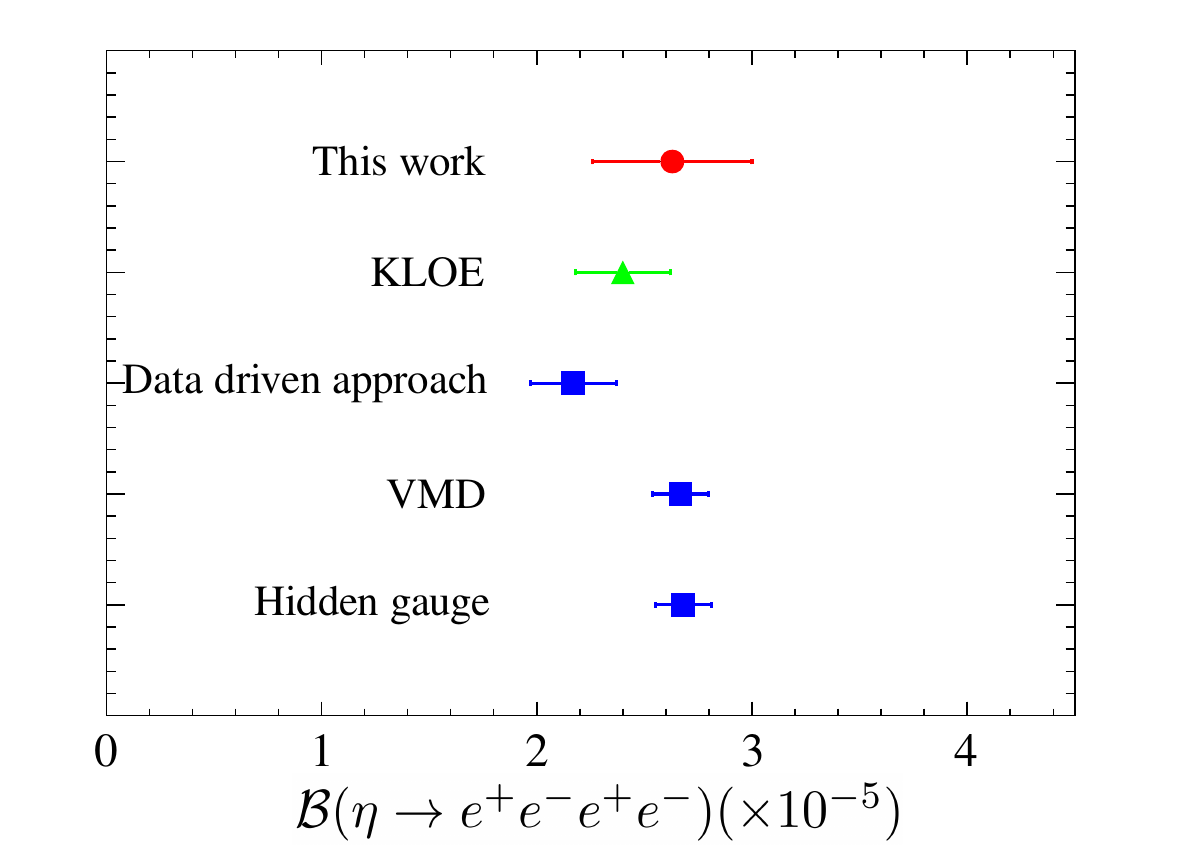}
\end{minipage}
\caption{Comparison of the branching fraction of $\eta\rightarrow e^+e^-e^+e^-$ from this work and the KLOE experiment~\cite{KLOE:2011qwm}; results of theoretical calculations are also given~\cite{VMD,DDA}.}
\label{result}
\end{center}
\end{figure}

The BESIII Collaboration thanks the staff of BEPCII (https://cstr.cn/31109.02.BEPC) and the IHEP computing center for their strong support. This work is supported in part by National Key R\&D Program of China under Contracts No. 2025YFA1613900, No. 2023YFA1606000, No. 2023YFA1606704; 
National Natural Science Foundation of China (NSFC) under Contracts No. 12247101, No. 11635010, No. 11935015, No. 11935016, No. 11935018, No. 12025502, No. 12035009, No. 12035013, No. 12061131003, No. 12192260, No. 12192261, No. 12192262, No. 12192263, No. 12192264, No. 12192265, No. 12221005, No. 12225509, No. 12235017, No. 12361141819; 
the Fundamental Research Funds for the Central Universities No. lzujbky-2025-ytA05, No. lzujbky-2025-it06, No. lzujbky-2024-jdzx06; the Natural Science Foundation of Gansu Province No. 22JR5RA389, No. 25JRRA799; the ``111 Center'' under Grant No. B20063; 
the Chinese Academy of Sciences (CAS) Large-Scale Scientific Facility Program; the Strategic Priority Research Program of Chinese Academy of Sciences under Contract No. XDA0480600; CAS under Contract No. YSBR-101; 100 Talents Program of CAS; The Institute of Nuclear and Particle Physics (INPAC) and Shanghai Key Laboratory for Particle Physics and Cosmology; ERC under Contract No. 758462; German Research Foundation DFG under Contract No. FOR5327; Istituto Nazionale di Fisica Nucleare, Italy; Knut and Alice Wallenberg Foundation under Contracts No. 2021.0174, No. 2021.0299; Ministry of Development of Turkey under Contract No. DPT2006K-120470; National Research Foundation of Korea under Contract No. NRF-2022R1A2C1092335; National Science and Technology fund of Mongolia; Polish National Science Centre under Contract No. 2024/53/B/ST2/00975; STFC (United Kingdom); Swedish Research Council under Contract No. 2019.04595; U. S. Department of Energy under Contract No. DE-FG02-05ER41374.

\bibliographystyle{apsrev4-2}
\bibliography{myref}

\newpage
\begin{widetext}
\begin{center}
M.~Ablikim$^{1}$\BESIIIorcid{0000-0002-3935-619X},
M.~N.~Achasov$^{4,b}$\BESIIIorcid{0000-0002-9400-8622},
P.~Adlarson$^{81}$\BESIIIorcid{0000-0001-6280-3851},
X.~C.~Ai$^{86}$\BESIIIorcid{0000-0003-3856-2415},
R.~Aliberti$^{39}$\BESIIIorcid{0000-0003-3500-4012},
A.~Amoroso$^{80A,80C}$\BESIIIorcid{0000-0002-3095-8610},
Q.~An$^{77,64,\dagger}$,
Y.~Bai$^{62}$\BESIIIorcid{0000-0001-6593-5665},
O.~Bakina$^{40}$\BESIIIorcid{0009-0005-0719-7461},
Y.~Ban$^{50,g}$\BESIIIorcid{0000-0002-1912-0374}
% ...
\\
\vspace{0.2cm}
(BESIII Collaboration)\\
\vspace{0.2cm} {\it
$^{1}$ Institute of High Energy Physics, Beijing 100049, People's Republic of China\\
% ...
$^{4}$ Budker Institute of Nuclear Physics SB RAS (BINP), Novosibirsk 630090, Russia\\
% ...
$^{39}$ Johannes Gutenberg University of Mainz, Johann-Joachim-Becher-Weg 45, D-55099 Mainz, Germany\\
$^{40}$ Joint Institute for Nuclear Research, 141980 Dubna, Moscow region, Russia\\
% ...
$^{50}$ Peking University, Beijing 100871, People's Republic of China\\
% ...
$^{62}$ Southeast University, Nanjing 211100, People's Republic of China\\
% ...
$^{64}$ State Key Laboratory of Particle Detection and Electronics, Beijing 100049, Hefei 230026, People's Republic of China\\
% ...
$^{77}$ University of Science and Technology of China, Hefei 230026, People's Republic of China\\
% ...
$^{80}$ University of Turin and INFN, (A)University of Turin, I-10125, Turin, Italy; (B)University of Eastern Piedmont, I-15121, Alessandria, Italy; (C)INFN, I-10125, Turin, Italy\\
$^{81}$ Uppsala University, Box 516, SE-75120 Uppsala, Sweden\\
% ...
$^{86}$ Zhengzhou University, Zhengzhou 450001, People's Republic of China\\

\vspace{0.2cm}
$^{\dagger}$ Deceased\\
% ...
$^{b}$ Also at the Novosibirsk State University, Novosibirsk, 630090, Russia\\
% ...
$^{g}$ Also at State Key Laboratory of Nuclear Physics and Technology, Peking University, Beijing 100871, People's Republic of China\\
% ...

}

\end{center}
\end{widetext}

\end{document}